# Mini review

# Mechanisms promoting biodiversity in ecosystems

Ju Kang[1,†], Yiyuan Niu[1,†], Xin Wang[1,*]

[1]School of Physics, Sun Yat-sen University, Guangzhou 510275, China

[†]These authors contributed equally to this work
[*]Correspondence: wangxin36@mail.sysu.edu.cn

**Abstract**

Explaining biodiversity is a central focus in theoretical ecology. A significant obstacle arises from the Competitive Exclusion Principle (CEP), which states that two species competing for the same type of resources cannot coexist at constant population densities, or more generally, the number of consumer species cannot exceed that of resource species at steady states. The conflict between CEP and biodiversity is exemplified by the paradox of the plankton, where a few types of limiting resources support a plethora of plankton species. In this review, we introduce mechanisms proposed over the years for promoting biodiversity in ecosystems, with a special focus on those that alleviate the constraints imposed by the CEP, including mechanisms that challenge the CEP in well-mixed systems at a steady state or those that circumvent its limitations through contextual differences.

## Introduction

Biodiversity represents a fascinating aspect of life on Earth, encompassing both macroscopic and microscopic species inhabiting terrestrial and aquatic realms[1-4]. In tropical forest ecosystems, thousands of plant and animal species coexist[3], with a gram of soil harboring 2,000 or more microbial species[1]. In the sunlit zones of oceans, there are over 150,000 species of eukaryotic plankton[4]. However, explaining biodiversity is not a simple issue. In 2005, the question "What determined species diversity?" was listed in *Science* as one of 125 challenging scientific questions[2].

A significant obstacle in explaining biodiversity arises from the Competitive Exclusion Principle (CEP), also referred to as Gause's Law[5,6], which states that two species competing for a single resource cannot coexist at a steady state, as initially formulated by the theoretical foundation of Volterra[7] and the experimental studies of Gause[5]. In the 1960s, MacArthur and Levins extended the concept of CEP to ecosystems with an arbitrary number of resource species[8], stating that the number of consumer species cannot exceed that of resources at steady states. In fact, the concept of CEP was further abstracted in mathematical formulations, replacing the context of resource species with limiting factors[9,10]. However, for the clarity of discussion, definitions of CEP related to limiting factors are excluded in this review.

The conflict between CEP and biodiversity has been observed both in laboratories and in the wild. In nature, this conflict is exemplified by the famous paradox of the plankton, where in aquatic environments, a very limited number of abiotic resource species supports hundreds or more plankton species. In laboratories, Park reported in the 1950s that two different species of beetles coexisted for more than two years while competing for the same type of food (flour)[11]. Similarly, Ayala observed that two different species of flies cohabited for over 40 weeks while competing for the same abiotic resources in a bottle with serial transfer[12]. All these observations demonstrate violations of the CEP.

Over time, various mechanisms have been proposed to explain biodiversity in natural ecosystems. This review explores these mechanisms, focusing on those that alleviate the constraint of the CEP. Such mechanisms may either challenge the CEP in its original context, particularly in well-mixed and steady-state systems, or bypass its limitations through contextual differences. For clarity, our review is structured as follows: we will first introduce the classical proof of the CEP raised by MacArthur and Levins[8], along with the scenarios and formulations that would undoubtedly fall under the constraint of the CEP, and then examine mechanisms that transcend the limits set by the CEP.

## Classical Proof of the CEP

In the 1960s, MacArthur and Levin proposed a classical proof of the CEP for the generic case in which $S_C$ consumer species competing for $S_R$ resource species, wherein the CEP states that $S_C \leq S_R$ at steady states[8]. For convenience, first we rephrase the simplest case of two consumer species $C_1$ and $C_2$ competing for one resource species $R$ (i.e., $S_C=2$ and $S_R=1$, see Fig. 1A) where they considered the population dynamics of the system via a consumer-resource model in the following form:

$$\begin{cases} \dot{C}_i = C_i \left( f_i(R) - D_i \right), \quad i=1,2; \\ \dot{R} = g(R, C_1, C_2). \end{cases} \quad (1)$$

Here $f_i$ and $g$ represent unspecified functions, $D_i$ is the mortality rate of species $C_i$. At a steady state, if both consumer species coexist, it is expected that $f_i(R) = D_i$, where $i$=1, 2. This demands that the two curves $y = f_1(R)/D_1$ and $y = f_2(R)/D_2$ intersect the line $y=1$ at the same point, a situation typically unattainable (see Fig. 1B), unless specific constraints on model parameters are met (with Lebesgue measure zero). Consequently, two consumer species generally cannot coexist with one type of consumers at a steady state (see Fig. 1C). In the case of $S_C=3$ and $S_R=2$, a similar proof strategy can be employed (see Fig. 1D-F), and more broadly, this proof strategy can be applied to any cases with positive integer values of $S_C$ and $S_R$[8].

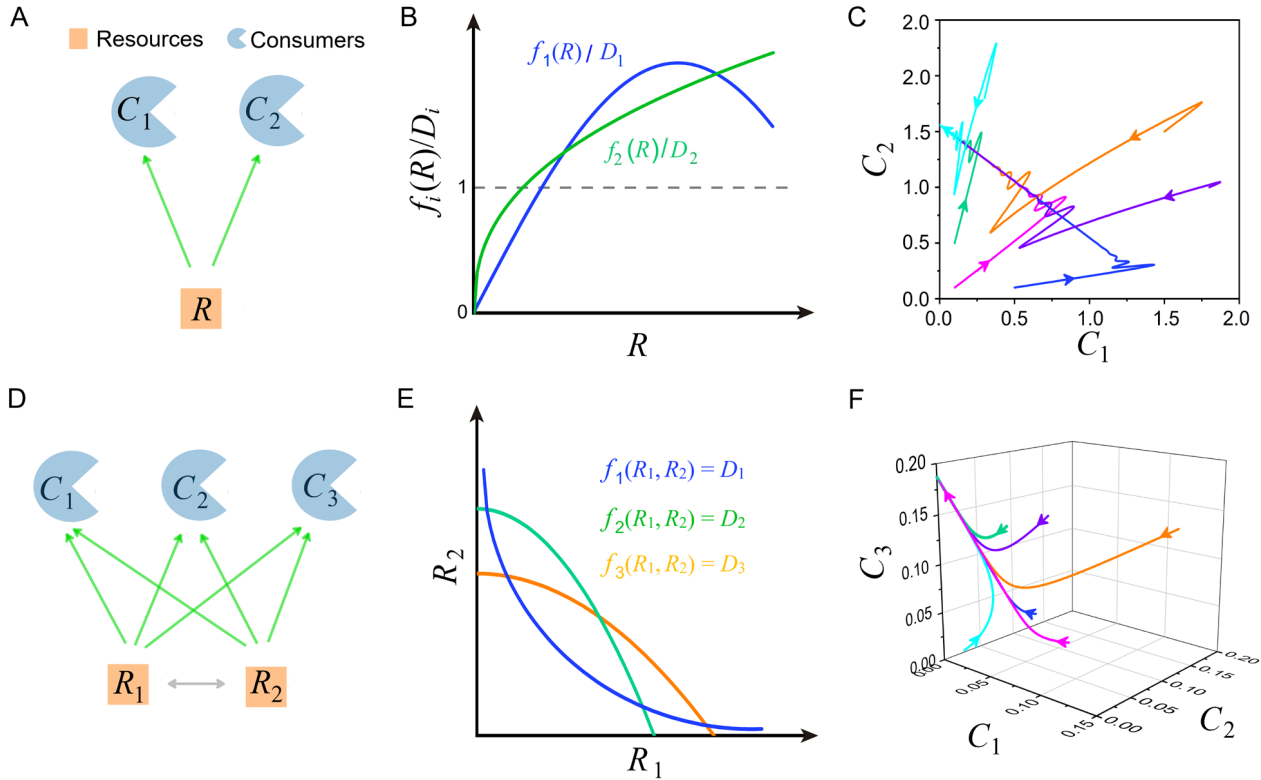

Fig. 1 | MacArthur and Levins' classical mathematical proof of the CEP[8]. (A, C) Cases where $S_C$ consumer species compete for $S_R$ resource species. The green arrows denote biomass flow between consumption relationships, with predation or facilitation forbidden among consumers but allowed among resources (illustrated by grey arrows). (B, E) At a steady state, the coexistence of all consumer species demands the intersection of three lines at the same point, which is typically impossible. (C, F) Phase trajectories of consumer species that cannot coexist in a steady state. Figure redrawn from the study of Wang and Liu[13].

## Scenarios or formulations under the constraint of CEP at a steady state

### Generalized Lotka-Volterra model is typically subject to the constraint of CEP

The pioneering work of Lotka[14] and Volterra[7] established the foundational mathematical models explaining species interactions, which were refined and expanded into the widely adopted Generalized Lotka-Volterra (GLV) model. Despite the extensive application of the GLV model[15-21], special attention needs to be paid to its application in tackling issues involving the CEP. This is due to the fact that freely chosen parameters in the GLV model implicitly imply subjection to the CEP[13].

The central idea could be illustrated using the simple case of the GLV model involving two types of competing species $C_i$ ($i=1,2$):

$$\begin{cases} \dot{C}_1 = C_1(\alpha_1 - \beta_{11}C_1 - \beta_{12}C_2), \\ \dot{C}_2 = C_2(\alpha_2 - \beta_{21}C_1 - \beta_{22}C_2), \end{cases} \quad (2)$$

where $\alpha_i$ and $\beta_{ij}$ $(i, j = 1, 2)$ are parameters. Generally, in GLV models, there is no specific constraint on the choice of parameters $\alpha_i, \beta_{ij}$. To illustrate the implicit subjection to the CEP, Wang and Liu[13] consider a consumer-resource model proposed by MacArthur[22] that comparable to this case with two consumer species and one abiotic resource species ($S_C$=2 and $S_R$=1):

$$\begin{cases} \dot{C}_1 = (\alpha_1' R - D_1)C_1, \\ \dot{C}_2 = (\alpha_2' R - D_2)C_2, \\ \dot{R} = \frac{r_a}{K_a} R\left[K_a(1 - R/K_a) - \beta_1' C_1 - \beta_2' C_2\right], \end{cases} \quad (3)$$

where $\alpha_i'$, $\beta_i'$, $D_i$ $(i, j = 1, 2)$, $r_a$, $K_a$ are parameters. In fact, by assuming a fast equilibrium condition for resource species, Eq. 3 can be reduced to Eq. 2, with $\alpha_i = \alpha_i' K_a - D_i$, $\beta_{ij} = \alpha_i' \beta_j'$ $(i, j = 1, 2)$. However, this leads to a strict constraint on parameters $\beta_{ij}$: $\frac{\beta_{11}}{\beta_{12}} = \frac{\beta_{21}}{\beta_{22}}$, severely restricting the choice of parameters in a GLV model. The analysis above is applicable to GLV models involving more consumer species, resulting in only when GLV models are subject to CEP can their parameters be freely chosen[13].

**The scenario involving only chasing pairs is subject to the constraint of CEP**

For well-mixed ecosystems, the population dynamics of the system can be rigorously derived from microscopic species interactions using Mean Field Theory borrowed from statistical physics. This method is widely used in studying chemical reactions and is suitable for ecological systems if the consumer species can move freely. By applying this method, Wang and Liu investigated whether the scenario involving only chasing pairs is subject to the constraint of CEP at steady state[13].

The model for the simple case of two consumer species competing for one resource species ($S_C$=2 and $S_R$=1) is shown in Fig. 2A[13], where the population structure of the consumers and resources is explicitly modeled: $C_i^{(F)}$ and $R^{(F)}$ represent the consumers and resources that are freely wandering (represented by the superscript "(F)"). When a consumer and a resource individual get close in space, the consumer chases the resource and they form a chasing pair $x_i = C_i^{(P)} \vee R^{(P)}$ (the superscript "(P)" represents pair). Then, the population abundances of consumer species $C_i$ and resource species $R$ are $C_i = C_i^{(F)} + x_i$ and $R = R^{(F)} + \sum_{i=1}^{2} x_i$, and the population dynamics is modeled as follows[13]:

$$\begin{cases} \dot{x}_i = a_i C_i^{(F)} R^{(F)} - (d_i + k_i) x_i, \\ \dot{C}_i = w_i k_i x_i - D_i C_i, \\ \dot{R} = g_1(R, C_1, C_2, x_1, x_2), \quad i = 1, 2, \end{cases} \quad (4)$$

where $a_i$, $d_i$, $k_i$, $w_i$, $D_i$, $R_a$, $K_a$ are model parameters, $g_1$ represents a unspecified function. At a steady state, if both consumer species coexist, it requires that $f_i(R^{(F)})/D_i = 1$ ($i$=1,2), where $f_i(R^{(F)}) = \dfrac{R^{(F)}}{R^{(F)} + K_i}$ and $K_i = \dfrac{d_i + k_i}{a_i}$ ($i$=1,2). This demands that two curves $y = f_1(R^{(F)})/D_1$ and $y = f_2(R^{(F)})/D_2$ intersect the line $y$=1 at the same point, rendering coexistence typically impossible (see Fig. 2C, E)[13].

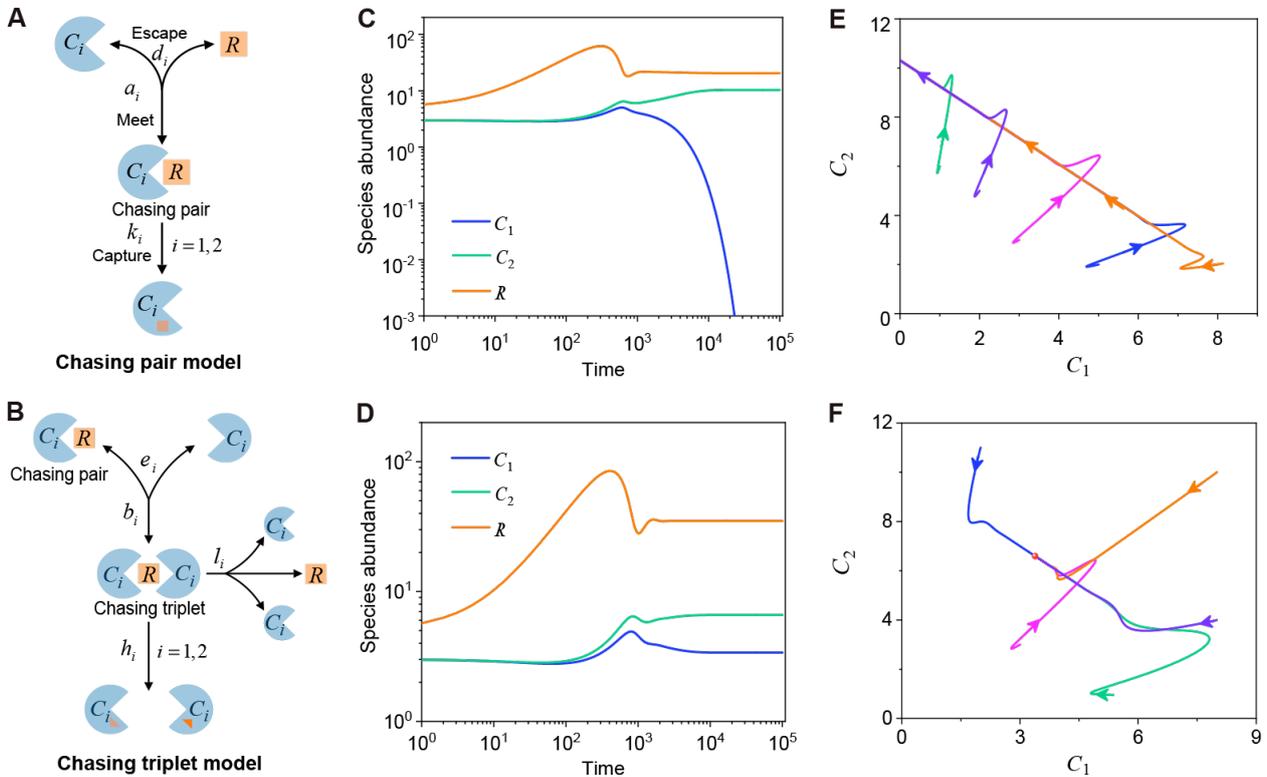

Fig. 2 | A scenario involving only chasing pairs is subject to the constraint of CEP, whereas further integration with chasing triplets can naturally break the CEP[13]. (A) Formation of chasing pairs within the consumption process between consumers and resource. (B) Formation of chasing triplet involving two consumers and a resource. (C, E) Consumers species cannot coexist at a steady state in the scenario involving only chasing pairs. (D, F) Coexistence of consumers in the scenario involving both chasing pairs and chasing triplets. Figure redrawn from the study of Wang and Liu[13]

## Mechanisms liberating the constraint of CEP through contextual differences

Below, we introduce mechanisms promoting biodiversity by liberating the constraint of CEP. First, we examine those achieving this through contextual differences.

**Temporal variations in the environment**

In 1961, in a famous work on the "Paradox of Plankton," Hutchinson proposed that temporal variation could be a crucial factor for bypassing the constraint of CEP in the wild[23]. Indeed, temporal variations are widespread in nature, such as day-night cycles, seasonal shifts, and unpredictable weather variations like alternations among precipitation, sunshine, wind, and snowfall (refer to Fig. 3A). Advantageous species could vary with changing environmental factors, and thus, if competition does not lead to species extinction before environmental changes occur, then equilibrium is never reached, allowing species to bypass the CEP due to this factor[23].

In the subsequent studies, Levins proposed that consumer populations not only exploit resources but also create new ecological niches amidst temporal variability and nonlinear dynamics[24]. This process may hinder ecosystems from attaining stable equilibrium. In 2006, Adler et al. conducted a meticulous examination of three decades of demographic data from a Kansas prairie, unveiling how interannual climate variability facilitates the coexistence of three dominant grass species[25]. This climate variability enhances growth rates at low densities, thereby preventing competitive exclusion.

**Spatial Heterogeneity**

Spatial heterogeneity is a common feature of ecological environments, presenting as a mosaic of diverse subenvironments inhabited by different species (see Fig. 3B), rather than adhering to a uniformly mixed, homogeneous condition. In 1974, Levin proposed that spatial heterogeneity contributes to increased species diversity[26]. This concept can be illustrated using the example depicted in Fig. 3B, where three consumer species ($C_1$, $C_2$ and $C_3$) compete for two resource species ($R_1$ and $R_2$) within an area comprised of three distinct patches. If each patch is occupied by a different consumer species, then, although each patch is subject to the constraint of the CEP, when considered as a whole, it is evident that there are more consumer species than resource species (Fig. 3B). Furthermore, Levin elucidated that heterogeneity can emerge in an initially homogeneous environment due to random initial events, such as colonization patterns, with their effects magnified by species interactions.

The influence of spatial heterogeneity is widely documented in studies of natural ecosystems. Within plant communities, the spatial arrangement of various species often exhibits gaps and patches[27,28]. In aquatic ecosystems like lakes, variations in factors such as light exposure and temperature can lead to chemical stratification across different depths[29]. For plankton communities, sampling studies in the 1970s at Castle Lake, California, revealed significant patchiness among various phytoplankton species[30]. In 2015, expanded research conducted during the Tara Oceans expedition collected ocean samples worldwide, revealing diversified planktonic communities and significant vertical stratification mainly driven by temperature differences in depth[4,31]. Recently, spatial heterogeneity incorporating ocean currents has been included in modeling studies[32,33].

**Self-organized dynamics**

For well-mixed systems in a stable environment, self-organized dynamics may naturally arise from species interactions, leading to species coexistence in an oscillatory or chaotic manner. This disrupts the attainment of an equilibrium state and bypasses the constraint of CEP (see Fig. 3C).

Specifically, in 1974, Koch proposed a model for the case involving two species of consumers and one species of biotic resources ($S_C$=2 and $S_R$=1), and demonstrated that in deterministic simulation studies, both consumer species could coexist through oscillations (Fig. 3C), naturally breaking the constraint of CEP by bypassing a steady state[34]. In 1999, Huisman and Weissing[35] conducted modeling studies for systems involving multiple resource species and demonstrated the enduring coexistence of many consumer species through oscillations and chaos in the population dynamics with deterministic simulation studies.

In experimental systems, evidence was reported for the relevance of promoting biodiversity through self-organized dynamics. For instance, in 2005, Becks et al. reported both oscillating and chaotic species coexistence for microbial species in chemostat systems[36]. In 2008, Benincà et al. reported experimental observations of chaos in the analysis of a 6-year long-term coexistence of a plankton community from the Baltic Sea and suggested chaotic behavior could be a plausible explanation for the paradox of plankton[37].

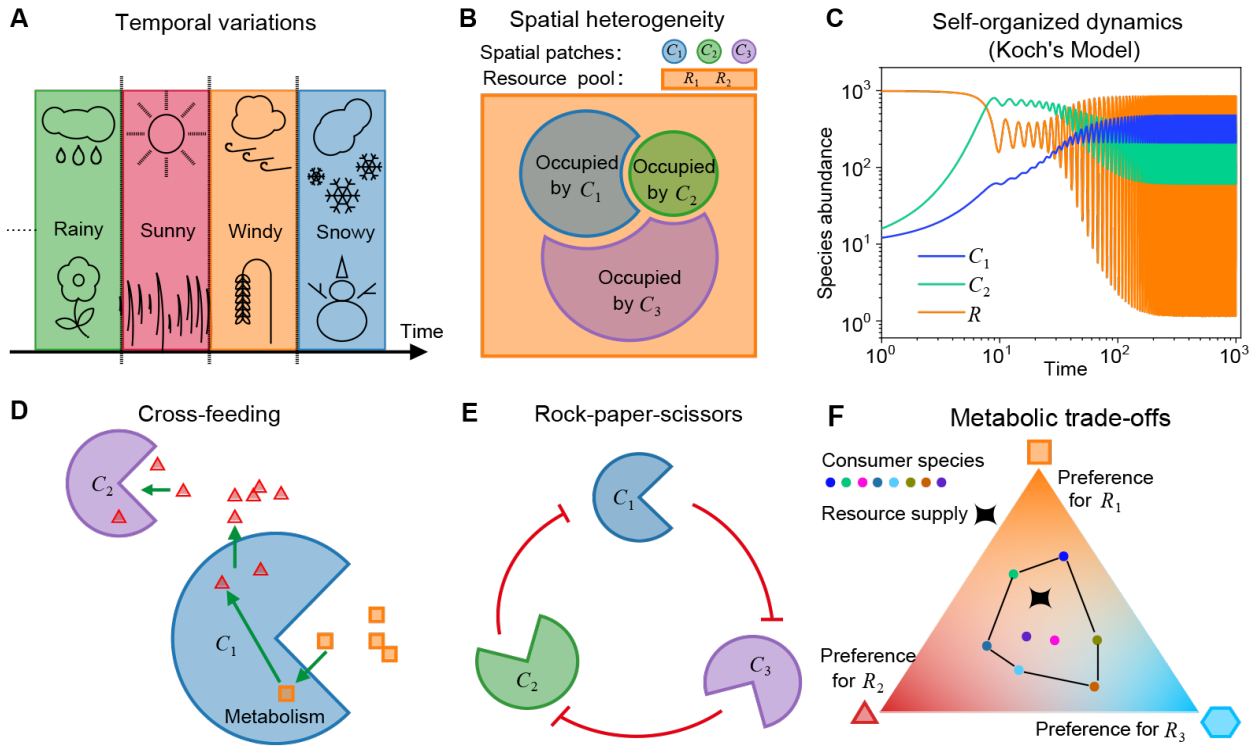

Fig. 3 | Mechanisms overcoming the constraint set by CEP through contextual differences. (A) Temporal variations in the environments[23,24]. (B) Spatial heterogeneity or patchiness[26]. (C) Self-organized dynamics such as oscillating coexistence[34]. (D) An illustration of cross-feeding among microbial species[38]. (E) An illustration of the rock-paper-scissors relationship among consumer species[39]. (F) Metabolic trade-offs in microbial communities. With different resource preferences, multiple consumers may coexist in stochastic simulation studies when the external resource supply lies within the convex hull of the consumers, a closed region delineated by black lines[40].

## Cross-feeding

Cross-feeding is a prevalent phenomenon in microbial ecosystems, where microbial species excrete metabolites that serve as nutrients into the culture medium or their surrounding environment. This process leads to the production of additional resources beyond those initially provided (see Fig. 3D). Consequently, if only the externally supplied resources are considered, microbial ecosystems may appear to violate the Competitive Exclusion Principle (CEP). However, when the resources secreted by microbes are taken into account, this mechanism operates within the constraints of the CEP.

Cross-feeding stands out as one of the most effective ways to promote biodiversity in microbial ecosystems. Goyal and Maslov employed model simulation studies to illustrate that microbial ecosystems exhibit high diversity, stability, and partial reproducibility even with only one type of resource, owing to cross-feeding[41]. In 2018, Goldford et al. reported experimental findings highlighting the widespread occurrence of nonspecific cross-feeding in microbial communities. They observed the stable coexistence of multiple microbial species with the external supply of just a single type of carbon source[38].

More recently, Lopez and Wingreen introduced a theory known as the noisy metabolism-averaging theory, aimed at elucidating the emergence of inter-species cross-feeding. Their theory is grounded in the premise that bacteria, owing to their small size, are vulnerable to the noisy regulation of metabolism, thereby constraining their growth rate. To counteract this limitation, closely related bacteria can engage in the sharing of metabolites to "average out" noise and bolster their collective growth[42].

**Complex interspecific interactions**

In the wild, species may coexist through complex interactions such as Rock-Paper-Scissors relations and higher-order interactions. The Rock-Paper-Scissors relation, observed in ecosystems where three species, $C_1$, $C_2$ and $C_3$, follow the rule where rock dominates scissors, scissors prevail over paper, and paper trumps rock (see Fig. 3E). This intransitive competition relationship ensures that species emerge victorious in at least one pairwise competition, thereby preventing the dominance of a single all-winning species and promoting coexistence[43].

The Rock-Paper-Scissors relation is commonly observed in bacterial communities because of the addition effects of antibiotic excretions and antibiotic resistance[39,44,45]. This relation has also been reported in macroscopic natural ecosystems, such as the cryptic coral reef system[46] and lizard communities[47]. Still, temporal environmental change and demographic factors may also play a role in affecting species competitiveness in the wild, which may thus shape the Rock-Paper-Scissors relation.

Higher-order interaction implies that the presence of a third species modifies the original interaction between two other species. For example, if a toxin produced by species $C_1$ inhibits species $C_2$, while species $C_3$ is capable of degrading this toxin, then it exerts a higher-order interaction. In 2015, Kelsic et al. studied the integration of high-order interaction with the Rock-Paper-Scissors relation in microbial communities and revealed a broad spectrum of coexistence patterns through modeling studies ranging from stable coexistence, oscillations to chaos, when multiple antibiotics are involved[48]. In 2016, Bairey et al. found that higher-order interaction can facilitate ecosystems incorporating a high level of species diversity to stably coexist[49]. This is validated by Grilli et al.'s theoretical studies[50] in which they revealed the stabilizing role of higher-order interaction, turning oscillatory and chaotic coexistence into stable coexistence, thereby increasing the vulnerability of ecosystems.

**Metabolic trade-offs**

Metabolic trade-offs refer to the constraints faced by consumer species due to a trade-off in resource allocation for feeding on different resource species[40]. In microbial communities, protein allocation is a notable constraint leading to metabolic trade-offs. In 2017, Posfai et al. introduced a pioneering model incorporating this mechanism, elucidating that with the introduction of stochasticity, multiple microbial species can coexist with three types of resources when the external supply of resources falls within a certain range of resource preference region set by the consumer species (see Fig. 3F)[40].

In practice, the deterministic version of Posfai et al.'s model[40] satisfies the functional form considered in MacArthur and Levin's proof of the CEP[8]. Meanwhile, stochasticity is prone to jeopardizes species coexistence in ecological models[51]. However, for the mechanism of metabolic trade-offs, stochasticity plays a role similar to that of stochastic resonance[52], facilitating species coexistence through the integration of metabolic trade-offs, thereby promoting biodiversity in microbial ecosystems.

## Mechanisms breaking CEP in its original context

Below, we review mechanisms for overcoming CEP in its original context, within the framework of well-mixed systems operating at a steady state.

### Chasing triplets

Pack hunting is a phenomenon commonly observed across different organisms[53-61], wherein multiple consumer individuals pursue a resource individual simultaneously. Building on these observations, Wang and Liu proposed a model that extended the consideration of chasing pairs to chasing triplets[13], where a consumer can join an existing chasing pair $x_i = C_i^{(P)} \vee R^{(P)}$ to form a chasing triplet $y_i = C_i^{(T)} \vee R^{(T)} \vee C_i^{(T)}$ (see Fig. 2B, the superscript "(T)" represents triplet). For a system containing two consumer species and one abiotic resource species ($S_C=2$ and $S_R=1$), the population abundances of consumers and resources are $C_i = C_i^{(F)} + x_i + 2y_i$ and $R = R^{(F)} + \sum_{i=1}^{2}(x_i + y_i)$ $(i=1,2)$, and the population dynamics is modeled as follows[13]:

$$\begin{cases} \dot{x}_i = a_i C_i^{(F)} R^{(F)} - (d_i + k_i) x_i - b_i x_i C_i^{(F)} + e_i y_i, \\ \dot{y}_i = b_i x_i C_i^{(F)} - (h_i + e_i + l_i) y_i \\ \dot{C}_i = w_i (k_i x_i + h_i y_i) - D_i C_i, \\ \dot{R} = R_a(1 - R/K_a) - \sum_{i=1}^{2} k_i x_i - \sum_{i=1}^{2} h_i y_i, \quad i=1,2. \end{cases} \quad (5)$$

where $a_i, b_i, d_i, e_i, h_i, k_i, l_i, w_i, D_i, R_a, K_a$ are model parameters. At a steady state, the zero-growth isoclines of the three species (i.e., $\dot{C}_1 = 0$, $\dot{C}_2 = 0$ and $\dot{R} = 0$) correspond to non-parallel surfaces in the ($C_1, C_2, R$) coordinates, which shares a common point (see Fig. 5 E). Such a fixed point can be stable, and hence the mechanism of chasing triplets enables two consumer species to coexist with one type of resources at steady state (see Fig. 2D, F), thereby breaking CEP in its original context[13].

### Intraspecific predator interference

Predator interference refers to the pairwise encounters between consumer individuals, encompassing interactions ranging from subtle staring contests to overt physical confrontations. In 1975, both Beddington[62] and DeAngelis et al.[63] independently introduced identical phenomenological models to elucidate the influence of predator interference on the functional response of consumer species. These models were later collectively referred to as the Beddington-DeAngelis (B-D) model by subsequent studies. The functional response of the B-D model is represented as:

$$\mathcal{F}(R, C_i) = \frac{aR}{1 + at_h R + a't_w C_i},\tag{6}$$

where $a$, $a'$, $t_h$ and $t_w$ are model parameters. Note that the functional response described above (Eq. 6) includes consumer dependency and thus does not conform to the functional form outlined in MacArthur and Levins' classical proof[8]. Consequently, the B-D model has been used in several studies to liberate the constraint of CEP[64,65]. However, it's worth noting that the functional response of the B-D model can be derived from scenarios involving only chasing pairs without predator interference[13,66]. This casts doubt on the appropriateness of employing the B-D model to break the CEP, as scenarios involving only chasing pairs are subject to the constraints of CEP[13].

Most recently, Kang et al. developed a mechanistic model of predator interference to address this issue;[67], wherein they applied Mean Field Theory to model the pairwise encounters between consumer and resource individuals (see Fig. 4A). For a system containing $S_C$ consumer species and $S_R$ abiotic resource species, considering a scenario involving chasing pairs and both intra- and interspecific predator interference, the population dynamics are modeled as follows:

$$\begin{cases} \dot{x}_{il} = a_{il} C_i^{(F)} R_l^{(F)} - (d_{il} + k_{il}) x_{il}, \\ \dot{z}_{ii} = a'_{ii} \left[ C_i^{(F)} \right]^2 - d'_{ii} z_{ii}, \\ \dot{z}_{ij} = a'_{ij} C_i^{(F)} C_j^{(F)} - d'_{ij} z_{ij}, \quad i \neq j, \\ \dot{C}_i = \sum_{l=1}^{S_R} w_{il} k_{il} x_{il} - D_i C_i, \\ \dot{R}_l = R_a^{(l)} (1 - R/K_a^{(l)}) - \sum_{i=1}^{S_C} k_{il} x_{il}, \quad i = 1, \cdots S_C, l = 1, \cdots S_R, \end{cases} \tag{7}$$

where $x_{il} = C_i^{(P)} \vee R_l^{(P)}$ represents a chasing pairs, and $z_{ij} = C_i^{(P)} \vee C_j^{(P)}$ stands for a predator interference pair (with $i=j$ for intraspecific and $i \neq j$ for interspecific), $C_i = C_i^{(F)} + \sum_l x_{il} + 2 z_{ii} + \sum_{j \neq i} z_{ij}$ and $R_l = R_l^{(F)} + \sum_i x_{il}$, while $a_{il}$, $d_{il}$, $k_{il}$, $a'_{ij}$, $d'_{ij}$, $w_{il}$, $D_i$, $R_a^{(l)}$, and $K_a^{(l)}$ are model parameters. Using this model, Kang et al. demonstrated that with the mechanism of intraspecific predator interference, two or more, even hundreds of consumer species can coexist at steady state when there is only one type of abiotic resource species (see Fig. 4B-E). Furthermore, such coexistence states are robust to stochasticity (see Fig. 4B, C, E), and this model[67] can even quantitatively illustrate the rank-abundance curves observed in diversified ecological communities (see Fig. 4F), including communities of plankton[68,69], birds[70-72], bats[73] and many other organisms in the wild.

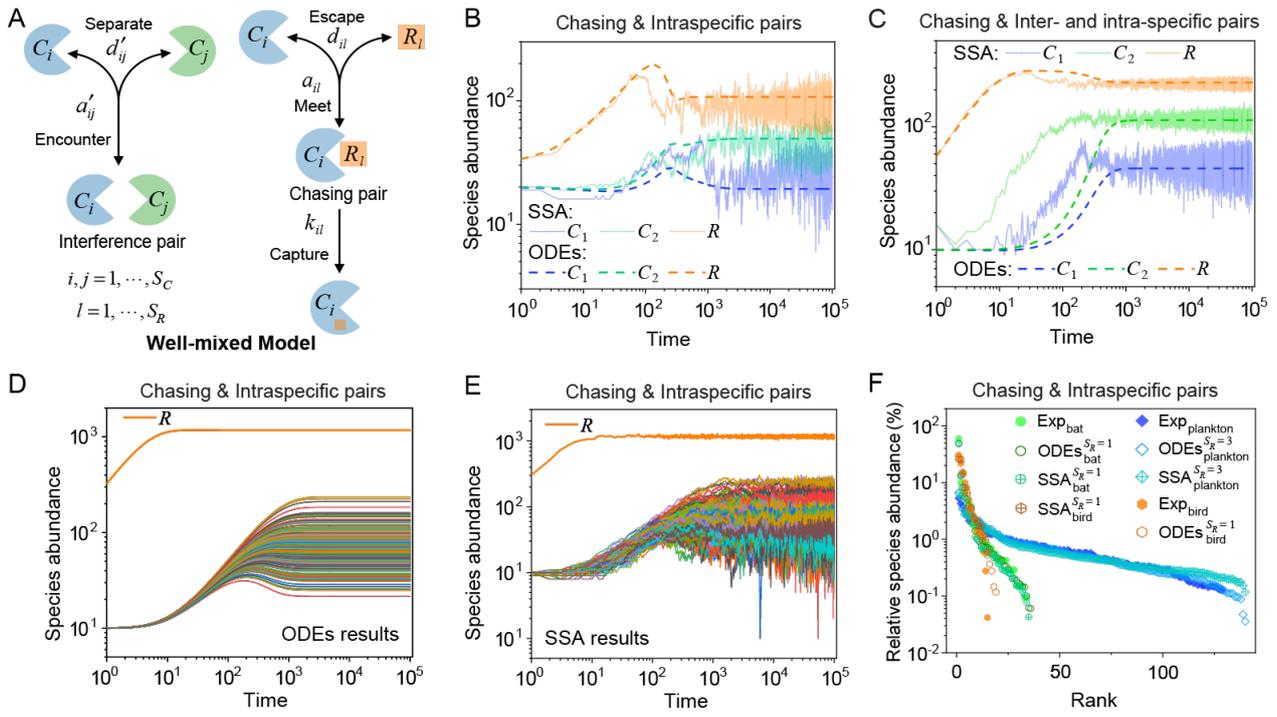

Fig. 4 | A mechanistic model of intraspecific predator interference breaks CEP and promotes biodiversity[67]. (A) Kang et al.'s mechanistic model of pairwise encounters. (B-C) Intraspecific predator interference facilitates the break of CEP. (D-E) Intraspecific interference enables multiple consumer species to coexist with one type of resource species irrespective of stochasticity. (F) Kang et al.'s model of intraspecific predator interference quantitatively illustrates the species' rank-abundance curves across diverse ecological communities. Figure redrawn from the study of Kang et al.[67].

**A necessary condition for mechanisms breaking CEP in its original context**

Intuitively, there is a quick test to check if a mechanism can break CEP in its original context. Consider the simplest case involving two consumer species $C_1$ and $C_2$, and one resource species $R$ ($S_C$=2 and $S_R$=1). The zero-growth isoclines of the three species, i.e., $\dot{C}_1 = 0$, $\dot{C}_2 = 0$ and $\dot{R} = 0$, correspond to three planes or surfaces in the ($C_1$, $C_2$, $R$) coordinates. A necessary condition for a mechanism to break CEP in its original context is that all these planes or surfaces are non-parallel to each other.

For instance, in MacArthur and Levins' classical proof of the CEP[8], if all species coexist, the zero-growth isoclines of species $C_1$ and $C_2$ correspond to two parallel planes governed by $f_i(R)/D_i = 1$ ($i$=1,2, see Eq. 1 and Fig.5A), which do not share a common point. Thus, a system described by Eq. 1 is subject to the constraint of CEP. In a scenario involving only chasing pairs, the zero-growth isoclines of species $C_1$ and $C_2$ correspond to two parallel surfaces governed by $f_i(R^{(F)})/D_i = 1$ ($i$=1,2, see Eq. 4 and Fig.5B, D). Therefore, this scenario is also under the constraint of CEP[13]. For scenarios involving chasing triplets or intraspecific predator interference, the zero-growth isoclines of the three species correspond to three non-parallel surfaces and thus share a common point (see Eqs. 5, 7, and Fig.5C, E, F). For these two mechanisms, the fixed points are stable, and thus both mechanisms[13,67] can break CEP at steady state in the original context.

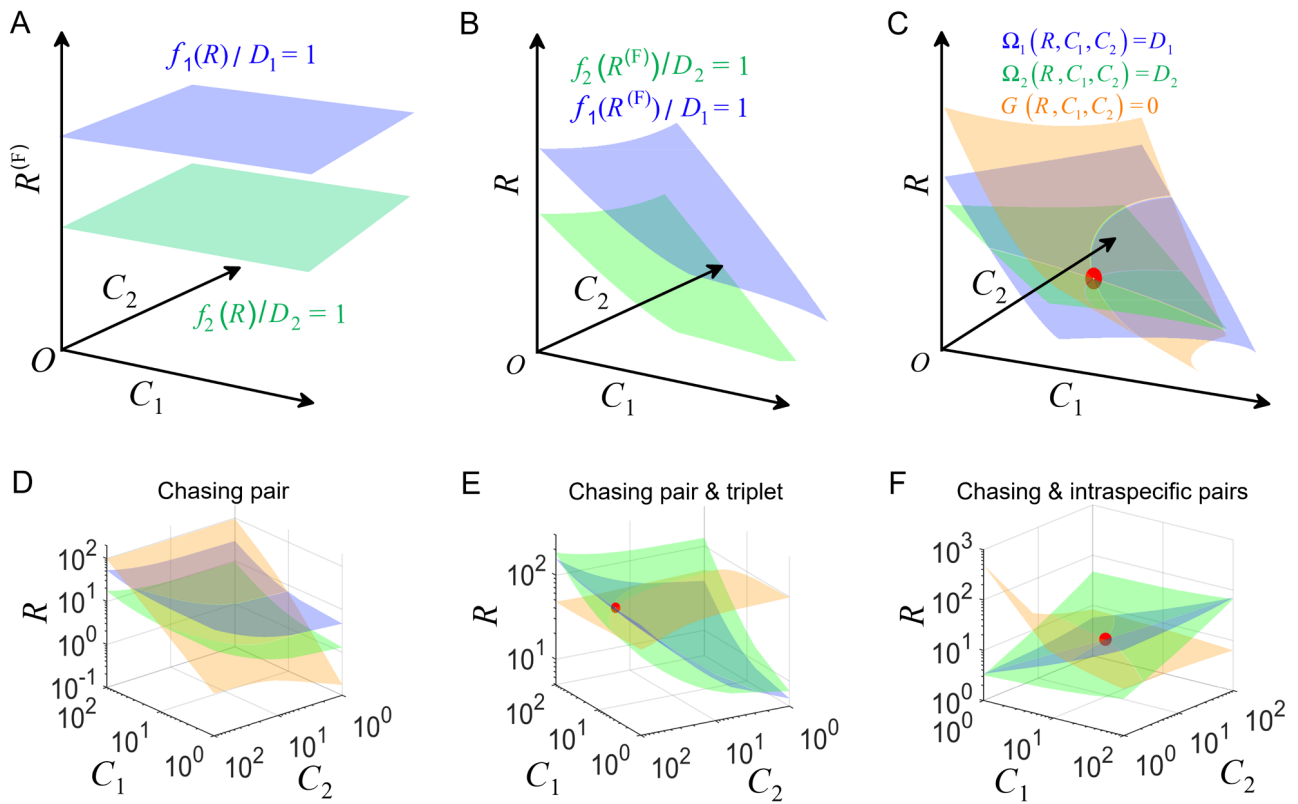

Fig. 5 | Intuitive understanding of mechanisms breaking CEP in its original context. Here, the blue surface, green surface, and orange surface represent the zero-growth isoclines of species $R$, $C_1$ and $C_2$, respectively. (A) In MacArthur and Levins' classical proof[8], the isoclines of species $C_1$ and $C_2$ correspond to two parallel planes. (B, D) In chasing pair model[13], the isoclines of species $C_1$ and $C_2$ correspond to two parallel surfaces. (C) A necessary condition for mechanisms breaking CEP at steady state is that the three surfaces have a common point. (E, F) Examples of the chasing triplet model[13], or mechanistic model of intraspecific predator interference[67], the fixed points shown in red are stable. Figure redrawn from the studies of Wang and Liu[13] and Kang et al.[67].

## Concluding remarks

The challenge posed by the CEP to biodiversity, as highlighted by the paradox of the plankton, has attracted intense research interest over the past several decades. In this review, we mainly introduce mechanisms that promote biodiversity by alleviating the constraints imposed by the CEP. Generally, these mechanisms can be classified into two different categories based on whether they adhere to the original context of the CEP.

The first category of these mechanisms breaks the constraint set by the CEP through circumventing its limitations via contextual differences. Among these, the most notable are mechanisms that assert ecosystems in nature never reach a steady state due to temporal variations[23,24,74], spatial heterogeneity[26,30], or self-organized dynamics such as oscillating and chaotic coexistence[34,35,37]. The first category also includes mechanisms such as cross-feeding, which operates within the constraint of the CEP if all secreted resources are counted, yet it plays an important role in maintaining species diversity, especially in microbial ecosystems. Metabolic trade-off[40] is also a mechanism that may contribute to biodiversity in microbial ecosystems, based on its validity in promoting biodiversity when stochasticity is introduced.

The second category of mechanisms breaks the CEP in its original context involving well-mixed systems and steady states. This includes mechanisms such as chasing triplets[13] and intraspecific predator interference[62,63,67]. For this type of category, special attention needs to be paid since approximations in the functional response may lead to errors in breaking the CEP. As mentioned earlier, the functional response of the B-D model[62,63] involving intraspecific predator interference could also be derived from scenarios involving only chasing pairs[13,66]. Since a scenario involving only chasing pairs is under the constraint of the CEP[13], it would inevitably cast doubt on the appropriateness of applying the B-D model[62,63] to break the CEP. Fortunately, Kang et al.'s study resolved this issue and proved that intraspecific predator interference could truly break the CEP and potentially resolve the paradox of the plankton based on rigorous examination[67]. Ultimately, a necessary condition for a mechanism to break the CEP while adhering to the original context requires that in the case of two consumer species competing for one resource species, the zero-growth isoclines of the three species should correspond to three non-parallel surfaces which share a common point (see Fig. 5C, E, F), and once the fixed point is stable, there would be a rigorous breaking of CEP in its original context.

Beyond mechanisms promoting biodiversity by alleviating the constraints set by the CEP, there are also many other mechanisms promoting species diversity through stabilizing species coexistence, such as higher-order interactions and rock-paper-scissors relations, where secreted metabolites or toxins are commonly involved in species interactions. For promoting biodiversity in natural ecosystems, many of the mechanisms mentioned above are potentially contributing, sometimes overlapping. It remains a challenge to identify the leading factors that promote biodiversity for specific real ecosystems.

## Acknowledgements

This work was supported by National Natural Science Foundation of China (Grant No.12004443), Guangzhou Municipal Innovation Fund (Grant No.202102020284) and the Hundred Talents Program of Sun Yat-sen University.

## Competing interests

The authors declare that they have no conflicts of interest.